\documentstyle[aps,preprint]{revtex}
\def\laq{\ \raise 0.4ex\hbox{$<$}\kern -0.8em\lower 0.62
ex\hbox{$\sim$}\ }
\def\gaq{\ \raise 0.4ex\hbox{$>$}\kern -0.7em\lower 0.62
ex\hbox{$\sim$}\ }

\begin{document}

%\preprint{\vbox{\baselineskip=12pt
%\rightline{BGU-PH-99/01}
%\vskip1truecm}}

\title{Causal Boundary Entropy From Horizon Conformal Field Theory }

\author{Ram Brustein}
\address{Department of Physics,
Ben-Gurion University, Beer-Sheva 84105, Israel\\ email:
ramyb@bgumail.bgu.ac.il}

\maketitle

\begin{abstract}

The quantum theory of near horizon regions of spacetimes with classical
spatially flat, homogeneous and isotropic Friedman-Robertson-Walker
geometry can be approximately described by a two dimensional conformal
field theory.  The central charge of this theory and expectation value of
its Hamiltonian are both proportional to the horizon area in units of
Newton's constant.  The statistical entropy of horizon states, which can be
calculated using two dimensional state counting methods, is proportional to
the horizon area and depends on a numerical constant of order unity which
is determined by Planck scale physics. This constant can be fixed such that
the entropy is equal to a quarter of the horizon area in units of Newton's
constant, in agreement with thermodynamic considerations. 

 \end{abstract}

Black holes possess geometric entropy equal to a quarter of the area of
their horizon in units of Newton's constant, known as the
Bekenstein-Hawking entropy \cite{Bekenstein:1973ur,Hawking:1974rv}. Since
the discovery of black hole entropy, many attempts were made to identify
its microscopic, statistical mechanics origin. Strominger
\cite{Strominger}, and more recently Carlip \cite{Carlip} have argued that
the statistical origin of black hole entropy is the ensemble of states of a
conformal field theory (CFT) describing fluctuations of two dimensional
(2D) horizon surfaces. They have used 2D methods to evaluate the density of
states of this theory, showing that their entropy is indeed a quarter of
the horizon area.  Some attempts were made to 
identify the horizon CFT \cite{Solodukhin}, 
and to extend the results to cosmological de Sitter space \cite{Wu}. 

It is widely accepted that geometric entropy must also be attributed to the
horizon of de Sitter space \cite{Gibbons}. The argument is that de Sitter
horizons are event horizons (as are black hole horizons), and therefore a
thermodynamic system crossing them is forever removed from an observer's
ken. Therefore the loss of the system's entropy must be compensated by an
increase of geometric entropy in order for the second law to remain valid.
In general, a cosmological horizon is not an event horizon, and a system
crossing it is not necessarily forever out of view and so, it may be
argued, there is no compelling reason to associate an entropy with a
cosmological horizon. 

In \cite{gsl}, I have proposed that geometric entropy has to be attributed
to cosmological horizons (or, in general, to causal boundaries), whether or
not they are event horizons. I have argued that entropy of quantum
fluctuations can be lost if the scale of causal connection becomes smaller
than their wavelength, for example, in an inflating universe. This is in
violation of the second law. The role of proposed geometric entropy is
precisely to restore validity of the second law in such situations. 

Here I show that the effective quantum theory of near horizon (NH)  regions
of spacetimes with classical spatially flat Friedman-Robertson-Walker (FRW) 
geometry is a 2D CFT, appearing due to huge redshifts suffered by horizon
fluctuations which allow only massless fluctuations to survive. The central
charge of this CFT $c=\alpha\frac{A^H}{4 \pi G_N}$, is proportional to the
horizon area $A^H$ in units of Newton's constant $G_N$, and depends on a
numerical constant of order unity $\alpha$, which is determined by Planck
scale physics.  The expectation value of the Hamiltonian of the theory
$L_0=\frac{A^H}{8 \pi G_N}$, is also proportional to the horizon's area in
units of Newton's constant. The asymptotic density of horizon states, and
therefore the horizon entropy $S^H$, can be obtained using Cardy's formula
\cite{Cardy} $S^H= 2\pi \sqrt{\frac{c}{6}\left(L_0-\frac{c}{24}\right)}$
(see \cite{Carlip} for detailed considerations about application of Cardy's
formula in this context). The cutoff dependent numerical coefficient
$\alpha$ can be set to $\alpha=6$, such that $S^H=\frac{A^H}{4 G_N}$, in
agreement with thermodynamic considerations. 

Horizons of FRW spaces are not necessarily event horizons, so our
results indicate that it is the existence of causal boundary which is the
source of geometric entropy, and that for a causal boundary to have
geometric entropy it is not required to hold information forever. Our
results strongly support the conjecture that causal boundaries and not only
event horizons have geometric entropies proportional to their area, and
therefore strengthen considerably the conclusion based on this conjecture
that a certain class of singularities are thermodynamically
forbidden \cite{gsl} (see also \cite{bek}).

Our starting point is the 4D Einstein-Hilbert action
\begin{equation}
 S^{(4)}=\frac{1}{16\pi G_N} \int d^4 x \sqrt{-g} R^{(4)}+S_m,
 \label{EHact}
\end{equation}
$G_N$ being Newton's constant, $g$ the determinant of the 4D metric
$g_{\mu\nu}$, $R^{(4)}$ is the 4D Ricci scalar, and $S_m$ is the matter
action. We consider spatially flat FRW solutions $ds_4^2= -dt^2 + a^2(t)
dr^2+ a^2(t) r^2 \left( d\theta^2+\sin^2\theta d\varphi^2\right)$, with
expanding scale factors
 $a(t)= a(t_0)
 \left(\frac{t}{t_0}\right)^\beta
 $.
The matter has an ideal fluid type energy momentum tensor derived from
$S_m$, given by $T^\mu_\nu=diag(\rho,-p,-p,-p)$. The energy density $\rho$,
and pressure $p$ are related by a simple equation of state $p= w \rho$,
which determines the scale factor expansion rate $\beta=\frac{2}{3(1+w)}$.
de Sitter space, for which the scale factor expands exponentially, should
be considered as the limiting case $w\rightarrow -1$, $\beta\rightarrow
\infty$. 

FRW spaces have finite causal connection scale $R_{CC}$ \cite{rbgv},
generically called ``horizon". Beyond the horizon local interactions are
not effective. This causal connection scale is determined by the Hubble
parameter $H=\frac{\dot a}{a}$, and its derivative: $R_{CC}^{-1} =\sqrt{
{\rm Max} \left[ \dot{H} + 2 H^2~, ~ - \dot{H}~\right] }$. In \cite{rbgv},
a covariant expression for $R_{CC}$ is given, but we will approximate it
here simply by $R_{CC}=H^{-1}$, a form applicable in almost all situations.
For $w<-\frac{1}{3}$ the expansion is inflationary, so fixed comoving
points ``inside the horizon", will in time ``exit the horizon", while for
$w>-\frac{1}{3}$ the expansion is decelerated, so fixed comoving points
``outside the horizon" , will in time ``enter the horizon".

Horizons are 2D surfaces that are classically, in the homogeneous and
isotropic cases that we are interested in here, well defined spherical
shells, which may evolve in time. Quantum mechanically they can fluctuate,
so it is reasonable to expect that they can be described effectively by 2D
field theories. 

We focus on the NH geometry by
choosing a 2D metric $ds_2^2= -dt^2 + a^2(t) dr^2$, fixing the
position of the horizon $H^{-1}(t)$, at some specified time $t^*$,
$H^{-1}(t^*)=d$, and changing coordinates to Schwartzschild-like
coordinates $R=a(t)\ r$, and $T$ defined by $dT= dt +
\frac{R/d}{1-R/d}\ dR$. In the new coordinates
\begin{eqnarray}
&& ds_2^2= -(1-H^2 R^2) dT^2\!
  -2\!\left[ HR- \frac{R}{ d}
 \frac{ 1- H^2 R^2 }{1-\left(\frac{R}{d}\right)^2} \right] dR dT
\nonumber \\ &&
 + \left[
 1 - \left(\frac{R}{d}\right)^2
 \frac{1- H^2 R^2} {\left(1-\left(\frac{R}{d}\right)^2 \right)^2}
 + 2HR\ \frac{R/d}{1-\left(\frac{R}{d}\right)^2}
  \right] dR^2\!\!.
\label{s1nh}
\end{eqnarray}
In the NH ``shell" defined by
 \begin{equation}
\frac{H^{-1}(t)-d}{d}\ll \frac{R-d}{d}\ll 1,
 \label{nhshell}
 \end{equation}
 (\ref{s1nh}) reduces to
\begin{equation}
ds^2_2\simeq -2 (1-R/d)\ dT^2 + \frac{1}{2} \frac{1}{1-R/d}\ dR^2.
\label{s2nh}
\end{equation}
One more change of coordinates
 $d\rho= \frac{1}{2}\frac{1}{1-R/d} dR$, brings
 (\ref{s2nh}) into
 $ds_2^2=  e^{-2\rho/d}\left(-dT^2 + d\rho^2\right)$
in the NH region $\rho/d \gg 1$, so that $\gamma_{ab}=
e^{-2\rho/d} \eta_{ab}$, $\sqrt{-\gamma}=e^{-2\rho/d}$, and
$\gamma^{ab}= e^{+2\rho/d} \eta_{ab}$. For later reference, we
note that $t$ is a NH lightcone coordinate,
\begin{equation}
t=T-\rho.
 \label{lc1}
\end{equation}

We would like to obtain an effective 2D field theory of NH
geometries, so  we parametrize the angular part of the metric
with a field $\Phi$, which will eventually determine the
horizon surface $\langle\Phi\rangle=H^{-1}(t)$,
 $
 ds_4^2= \gamma_{ab}(t,r) dx^a
dx^b+ \Phi^2(t,r,\theta,\phi) \left( d\theta^2+\sin^2\theta
d\varphi^2\right)
 $.  
$\Phi$ is allowed to have general dependence on the
coordinates. For this metric (\ref{EHact}) reduces to
\begin{equation}
  S^{(4)}= \frac{1}{8\pi G_N}  \int dt dr d\Omega_2 \sqrt{-\gamma}
  \Biggl\{
\gamma^{ab} \nabla_a\Phi \nabla_b\Phi + 
 \frac{1}{2} \Phi^2 R^{(2)} +1\Biggr\},
 \label{4Dact}
\end{equation}
where $d\Omega_2=\sin \theta d\theta d\phi$. In (\ref{4Dact}) we 
have dropped the matter action $S_m$, since its only function is to determine
the classical solution.
Note that terms containing angular 
derivatives are absent in (\ref{4Dact}) \cite{Yoon}.

Mass terms in action (\ref{4Dact}) are suppressed in the 
NH region (\ref{nhshell}) \cite{Solodukhin}. Obviously,  the last term in
(\ref{4Dact}) is exponentially suppressed by a factor
$e^{-2\rho/d}$ with respect to the first two terms\footnote{Strictly 
speaking, certain variations coming from this term are 
suppressed}.
The NH effective action is therefore the following,
\begin{equation}
 S^{(4)}_{\rm NH}= \frac{1}{8\pi G_N}  \int dT d\rho d\Omega_2
 \sqrt{-\gamma} 
 \Biggl\{  \gamma^{ab} \nabla_a\Phi \nabla_b\Phi +
 \frac{1}{2} R^{(2)} \Phi^2 \Biggr\}.
  \label{4Dnh}
\end{equation}

We proceed to reduce action (\ref{4Dnh}) to 2D. 
First, we expand $\Phi$ in spherical harmonics,
\begin{equation}
 \Phi(t,r,\theta,\phi)=
 \sum\limits^{l_{\rm max}}_{l,|m|\le l_{\rm max}}
 \Phi_{l,m}(t,r) Y_l^m(\theta,\varphi).
\label{expansion}
\end{equation}
The maximal angular
momentum $l_{\rm max}$ in expansion (\ref{expansion}) is
determined by the short distance cutoff of the theory, as we
discuss later. We then substitute (\ref{expansion}) into
(\ref{4Dnh}),  perform the angular integration  using the
orthogonality property of $Y_l^m$'s, and obtain a dimensionally
reduced  NH 2D effective action,
\begin{eqnarray}
 && S^{(2)}_{\rm NH}= \frac{1}{8\pi G_N}  \int dT d\rho
 \sqrt{-\gamma} \times
\nonumber \\ &&
 \Biggl\{ \sum_{l,|m|\le l_{\rm max}}^{l_{\rm max}}\!\!\!
 \gamma^{ab} \nabla_a\Phi_{l,m} \nabla_b\Phi_{l,m} +\!
 \frac{1}{2} R^{(2)} \!\!\!\! \sum_{l,|m|\le l_{\rm max}}^{l_{\rm max}}
 \!\!\! \Phi^2_{l,m}  \Biggr\}.
  \label{2Dnh}
\end{eqnarray}

Only a single field $\sum \Phi^2_{l,m}$, couples to the 2D
curvature term. 
This field is simply the area of the horizon shell $A^H$, 
$
 A^H = \int d\Omega_2\ \Phi^2(t,r,\theta,\phi)
 = \sum\limits^{l_{\rm max}}_{l,|m|\le l_{\rm max}}
 \Phi_{l,m}(t,r)^2.
$
Here we have performed the angular integration using the orthogonality
property of $Y_l^m$'s.  We will be interested in fluctuations of the
horizon which keep the area fixed at its (time-dependent) classical value,
and therefore we will freeze  quantum fluctuations of this mode.  
Since we are interested in counting states for the case of large
$l_{\rm max}$, projecting out a single mode will not compromise the
generality of our results. Freezing the quantum fluctuations of the
area has the benefit of simplifying quantization of the NH theory
enormously. Furthermore, we do not take into account fluctuations of matter
sources, assuming that their only effect is to determine the time dependent
expectation value of the horizon. 

So, when all is said and done, the remaining NH action
is simply a sum of actions of independent\footnote{We ignore the single
overall constraint on their sum, which effectively removes a single field.}
free scalar fields, minimally coupled to 2D gravity, 
\begin{equation}
S^{(2)}_{\rm NH}=\frac{1}{8\pi G_N}\int d^2x
 \sqrt{-\gamma}\!\!\!\sum_{l,|m|\le l_{\rm max}}^{l_{\rm
 max}}\hspace{-.2in}
 \gamma^{ab} \nabla_a\Phi_{l,m} \nabla_b\Phi_{l,m}.
\label{simple2Dnh}
\end{equation}
Theory  (\ref{simple2Dnh}) can be quantized in the 2D conformal
gauge, using standard  DDK arguments\cite{DDK,Polchinski}. The conformal anomaly
of $\Phi_{l,m}$'s induces a kinetic term for the 2D conformal mode, 
and renormalizes the 2D action such that  the full theory is a CFT, 
whose total central charge vanishes. 
But the central charge of the Liouville mode cannot be used to
calculate the density of states \cite{kutseib,Carlip}. 
For state counting purposes it counts as a single field.

To determine the horizon's entropy  we need to
compute the effective central charge $c$, of the NH CFT, 
and the expectation value of its
Hamiltonian $L_0$. We first calculate the total effective central charge $c$, 
which is approximately equal to the sum of individual
matter central charges,
\begin{equation}
 c\simeq\sum\limits_{l,|m|\le l_{\rm max}}^{l_{\rm max}}\hspace{-.1in}
 1\simeq l_{\rm max}^2.
  \label{ctot}
\end{equation}
In (\ref{ctot}) we have neglected contributions from ghosts, 
from the Liouville mode,  ignored the area constraint, and
included redundant contributions from a small number of gauge
modes, but since we are interested in the case of large 
$l_{\rm max}$, we are justified in doing so.

The maximal angular momentum $l_{\rm max}$, is determined by
Planck scale physics. The smallest angular variations
$\Delta\varphi_{\rm min}$ and $\Delta\theta_{\rm min}$ allowed as
fluctuations of a sphere of radius  $d$, are determined by the
short distance cutoff of the theory $\ell_{UV}$,
\begin{equation}
 \Delta\varphi_{\rm min}  = \ell_{UV}/d,  \hspace{.2in}
 \Delta\theta_{\rm min}  = \ell_{UV}/d.
\label{uv1}
\end{equation}
Since $Y_l^m\sim e^{im\varphi}e^{il\theta}$, the smallest angular
variations $\Delta\varphi_{\rm min}$ and $\Delta\theta_{\rm min}$
determine the maximal angular momentum,
\begin{equation}
 m_{\rm max} =\frac{C_m}{\Delta\varphi_{\rm min}},
\hspace{.2in}
 l_{\rm max} =\frac{C_l}{\Delta\theta_{\rm min}},
\label{uv2}
\end{equation}
where $C_l$, $C_m$ are numerical coefficients of order unity. We
may use eqs.~(\ref{uv1},\ref{uv2}) to estimate the maximal allowed
angular momentum $l_{\rm max}^2= C_l C_m\ d^2/\ell_{UV}^2$.
Assuming that the short distance cutoff is some numerical factor
of order unity $k$, times the Planck length $\ell_{UV}= k
\sqrt{G_N}$, and denoting $\alpha=\frac{C_l C_m}{k^2}$ we
obtain our final expression for the total central charge of the
NH theory,
\begin{equation}
c=\alpha \frac{A^H}{ 4\pi G_N}.
\label{ctotfin}
\end{equation}

The expectation value of $L_0$ is determined by the classical
background. Recall that the classical solution is a function of
time only $\langle\Phi(r,t,\theta,\varphi)\rangle=H^{-1}(t)$, and that
$H^{-1}(t)=d \left(\frac{t}{t^*}\right)$. But this means that only
the $l=0$, $m=0$ mode has non-trivial expectation value
\begin{equation}
 \langle\Phi_{0,0}\rangle=\sqrt{4\pi}d\left(\frac{t}{t^*}\right).
 \label{vev}
 \end{equation}
To simplify evaluation of $L_0$, we go to 2D lightcone coordinates
in two steps, first setting $\tau=\frac{T}{t^*}$,
$\sigma=\frac{\rho}{t^*}$, and then setting $x^\pm=\tau\pm\sigma$.
Note that according to (\ref{lc1}), $x^-=\frac{t}{t^*}$. We expand
\begin{equation}
\Phi_{0,0}=q+ p\ x^- + \sum_{n\ne 0} \frac{1}{n}\ \alpha_n\
e^{\hbox{$ 2 i n\pi\frac{\sigma}{\sigma_{\rm max}}\ x^-$}},
 \label{l0exp}
\end{equation}
where $\sigma_{\rm max}$ determines  the range of the 2D
coordinate $\sigma$. Since $x^-=\frac{t}{t^*}$, we can compare
(\ref{l0exp}) and (\ref{vev}), and observe that for the classical
background only $p$ is non-vanishing  $p= \sqrt{4\pi}d$, while all
the $\alpha_n$'s and $q$ vanish. Since 
$L_0=\frac{1}{8\pi G_N}\left[
p^2+\sum\limits_{n\ne 0} \alpha_n \alpha_{-n}\right]$, it follows
that $L_0=p^2/8\pi G_N$, but $p^2=A^H$, so
\begin{equation}
 L_0=\frac{A^H}{8\pi G_N}.
 \label{l0}
\end{equation}

In general, CFT's have two sets of independent modes which are either
functions of $x^+$, or $x^-$, but the NH theory has only
one set of modes \cite{Strominger,Carlip,Solodukhin}. 
As we show $\Phi_{l,m}=\Phi_{l,m}(x^-)$, leaving
only one Virasoro algebra as symmetry of the NH CFT.
Near the horizon, as we have already seen, propagating modes are
massless,  due to redshift effects. But the same redshift effects
allow them to propagate only along outgoing light-like
trajectories in the $x^-$ direction. Near black hole horizons,
similar redshift effects allow only
ingoing modes to propagate. To
see this in more detail, we look at the $x^+$ derivative
\begin{equation}
 \partial_+= \frac{1}{2} {\partial_T}_{|t}+
 \frac{1}{2}
{\partial_\rho}_{|t}
  = {\partial_T}_{| t} = {\partial_\rho}_{| t},
\end{equation}
where the last equation is obtained using eq.(\ref{lc1}). For
simplicity, we now set $t^*=1$, so $x^-=t$. Expressing
$T$ and $\rho$ derivatives in terms of $R$ derivatives, we find
\begin{equation}
  {\partial_\rho}_{| t} \simeq
  {\partial_T}_{| t} \simeq 2\left(1-\frac{R}{d}\right)\partial_R,
\end{equation}
so smooth functions in the original variables $(R,t)$ have
vanishing $x^+$ derivative in the NH region
(\ref{nhshell}). Further examination shows that all $x^+$
derivatives $\partial_+^n$, vanish for such functions,
 so  smooth functions in the original variables $(R,t)$
are not functions of $x^+$ but only functions of $x^-$, as
claimed\footnote{We will not discuss possible diffeomorphisms anomalies due
to the chiral nature of NH theory.}.

We are now ready to use Cardy's formula \cite{Cardy,Carlip} to
evaluate the entropy of the horizon,
\begin{eqnarray}
 S_{\rm horizon}&=& 2\pi
 \sqrt{\frac{c}{6}\left(L_0-\frac{c}{24}\right)}\nonumber \\
 &=& \frac{A^H}{4 G_N}\sqrt{\frac{\alpha}{3}
 \left(1-\frac{\alpha}{12}\right)},
\end{eqnarray}
where $\alpha$ is defined above (\ref{ctotfin}).

Since coefficient $\alpha$ is determined by short distance (Planck
scale) physics, we expect it to be universal. It should not be
sensitive to the macroscopic, large scale physics which determines
the exact nature of  classical solutions. If so, we may use the
limiting case of de Sitter space to ``calibrate" it. In de Sitter
space the horizon entropy is known to be $\frac{A^H}{4 G_N}$ from
other considerations. This procedure sets the value of $\alpha$ at
$\alpha=6$, leading to the conclusion that, in general,
\begin{equation}
 S_{\rm horizon}= \frac{A^H}{4 G_N}.
\end{equation}

The horizon shell, whose entropy we have just calculated, is a
thin shell, since its thickness is much smaller than
its radius, but its thickness is macroscopic, much larger than the
Planck length, so we might have  expected that its entropy turns
out to be proportional to its volume, in units of Planck volume.
The entropy of an arbitrary shell does scale as its volume, but
not the NH shell. So what is so special about the
horizon?  Huge redshifts ``squash"  NH
shells, and make their entropy proportional to their area and not to their
volume.

Our results apply also to collapsing pressureless matter, since the such
systems can be described by FRW metric. In this case,
the horizon whose entropy we have computed is an apparent horizon.
The apparent horizon reaches the event horizon when all matter has collapsed
to a point.

Our explicit calculations were carried out for 4-dimensional,
homogeneous and isotropic, spatially flat geometries. But the
essential ingredient in our calculation was the huge redshift near
the horizon. This left only massless outgoing 2D modes, which are
naturally described by a CFT. Since this ingredient seems to be
present whenever causal boundaries form,  I believe that similar
methods can be applied to higher dimensional spaces, along the
lines of \cite{Carlip,Solodukhin,Wu},  to more general spatially
curved spaces, using the covariant definition of causal connection
scale \cite{rbgv}, and to string theory along the lines of
\cite{bfs}.

\acknowledgements

I wish to thank S. de Alwis, S. Elitzur, S. Foffa and S. Solodukhin for
discussions, and CERN-TH division, where part of this work was
carried out, for hospitality.

\end{document}